\documentclass[final,1p,times]{elsarticle}

\usepackage{amssymb}
\usepackage{amsmath}

\newcommand{\gray}{${\gamma}$-ray\;}
\newcommand{\grays}{${\gamma}$-rays\;}

\newcommand{\ergcmsqs}{\mbox{erg~cm$^{-2}$~s$^{-1}$}}

\journal{JHEAP}

\journal{JHEAP}

\begin{document}

\begin{frontmatter}

\title{Multimessenger Emission from Very-High-Energy Black Hole-Jet Systems in the Milky Way}

\author[a]{Jose A. Carpio \corref{cor1}}
\ead{jose.carpiodumler@unlv.edu}

\author[a]{Ali Kheirandish}

\author[a]{and Bing Zhang}

\cortext[cor1]{Corresponding Author}

\affiliation[a]{Department of Physics & Astronomy; Nevada Center for Astrophysics, University of Nevada, Las Vegas, NV 89154, USA}

\begin{abstract}
Microquasars, compact binary systems with an accreting stellar-mass black hole or neutron star, are promising candidates for high-energy particle acceleration. Recently, the LHAASO collaboration reported on the detection of $>100$ TeV \gray emission from five microquasars, suggesting that these sources are efficient particle accelerators. In microquasars, high-energy $\gamma$-rays can be produced in large-scale jets or winds. In this work, we explore the X-ray, $\gamma$-ray and neutrino emission from SS 433, V4641 Sgr and GRS 1905+105. We consider leptonic and hadronic scenarios to explain the spectra observed by LHAASO and other high-energy $\gamma$-ray detectors. We estimate the neutrino flux associated with the hadronic component and investigate the detectability of neutrinos  from these sources in current and future neutrino telescopes. We find that among the three sources, V4641 Sgr has the best prospects of observation with a combined next-generation neutrino telescopes.  
\end{abstract}



\end{frontmatter}

\section{Introduction}

Microquasars are X-ray binaries with prominent jets, where a stellar-mass black hole pulls matter from a companion star, forming an accretion disk. The accretion disk potentially launches jets and winds, which offers particle acceleration sites ~\cite{Middleton:2018xiq}.
Microquasars in the Milky Way have been observed in different wavelengths and have historically been considered as potential targets for multimessenger observations with neutrinos~\cite{Gaisser:1983cj,Berezinsky:1985xp}. Further TeV $\gamma$-ray observations promoted neutrino flux predictions via either photohadronic or hadronuclear interactions in these sources~\cite{Aharonian:2005cx, Christiansen:2005gw,Torres:2006ub, Reynoso:2019vrp,Kimura:2020acy}.

Recently, the LHAASO collaboration reported the detection of very-high-energy (VHE) $E>100$ TeV \grays from 5 microquasars \cite{LHAASO:2024psv}. 
The detection of VHE \grays hints at efficient particle acceleration, suggesting that microquasars are good candidates for Galactic PeVatrons, provided that hadronic interactions are responsible for the VHE $\gamma$-rays. Microquasars have been considered as potential sites of acceleration for cosmic rays in the Milky Way (see e.g.,~\cite{Bosch-Ramon:2006cii} for an overview).
Potential particle acceleration mechanisms in these environments include diffusive shock acceleration~\cite{Drury:1983zz}, magnetic reconnection~\cite{PhysRevLett.108.135003}, and jet interactions with magnetically arrested disks \cite{Kuze:2025wda}. 

The most significant of the 5 LHAASO sources with identified VHE \gray is SS 433, with a reported significance of $12.9\sigma$. This source and two other sources, V4641 Sgr and GRS 1915+105, are extended \gray sources with angular extensions corresponding to sizes of the order of $10$ pc. For the other two sources, Cygnus X-1 and MAXI J1820+070, only upper limits on the \gray emission size have been reported~\cite{LHAASO:2024psv}. Therefore, the production site is likely more compact.  

The emission of TeV $\gamma$-rays can be attributed to a leptonic scenario, where accelerated electrons can up-scatter the cosmic microwave background (CMB) via inverse Compton (IC). In light of recent observations, the electrons would need to reach energies of $\sim$ 700 TeV in order to explain VHE $\gamma$-rays, which may be challenging in environments where strong synchrotron cooling is present. On the other hand, if PeV protons (or ions) are also accelerated, then VHE $\gamma$-rays can result from $\pi^0$ decays produced in hadronuclear ($pp$) collisions \cite{Romero2003,Bosch-Ramon:2008xrf}. For this scenario, a sufficiently high target proton density is also required for efficient $pp$ inelastic collisions. Alternatively, $p\gamma$ collisions may occur with ambient photons to enable photopion production. 

This hadronic explanation of $\gamma$-rays inevitably leads to an accompanying neutrino signal of a comparable flux, making high-energy neutrinos a powerful tool to investigate particle acceleration in microquasars. The multimessenger picture for neutrino and \gray emission has been considered for several microquasars such as SS 433, e.g., \cite{Reynoso:2019vrp,Kimura:2020acy}.

The IceCube Collaboration identified neutrino emission from the Galactic plane \cite{IceCube:2023ame}. 
The most significant observation is made for a quasi-diffuse neutrino emission based on the spectrum of the Fermi $\pi^0$ map. Lower significances were found for other templates as well as for the correlation studies aiming to identifying neutrinos from cataloged sources. 
At the moment, the nature of the observed emission from the Milky Way is not understood~\cite{Ambrosone:2023hsz}.  
Microquasars are potential contributors to the observed neutrino flux as well as Galactic cosmic rays. 
Particular interest is drawn to the microquasar contribution to the PeV flux around the knee of the cosmic ray spectrum (e.g.,~\cite{HESS:2024rlh,LHAASO:2024psv}).

In this work, we investigate the multimessenger emission from the $>$100 TeV extended regions identified by LHAASO coincident with microquasars SS 433, V4641 Sgr and GRS 1915+105. 
In Section II, we present our phenomenological model for the multimessenger emission. Next, we apply our model to the multi-wavelength data from these three sources and examine leptonic and hadronic hypotheses. For the hadronic models, we study the associated neutrino emission and observation prospects for current and future experiments. Finally, in Section IV we summarize our results.

\section{Method}\label{sec:method}

\begin{figure}
\centering
\includegraphics[width=0.8\columnwidth]{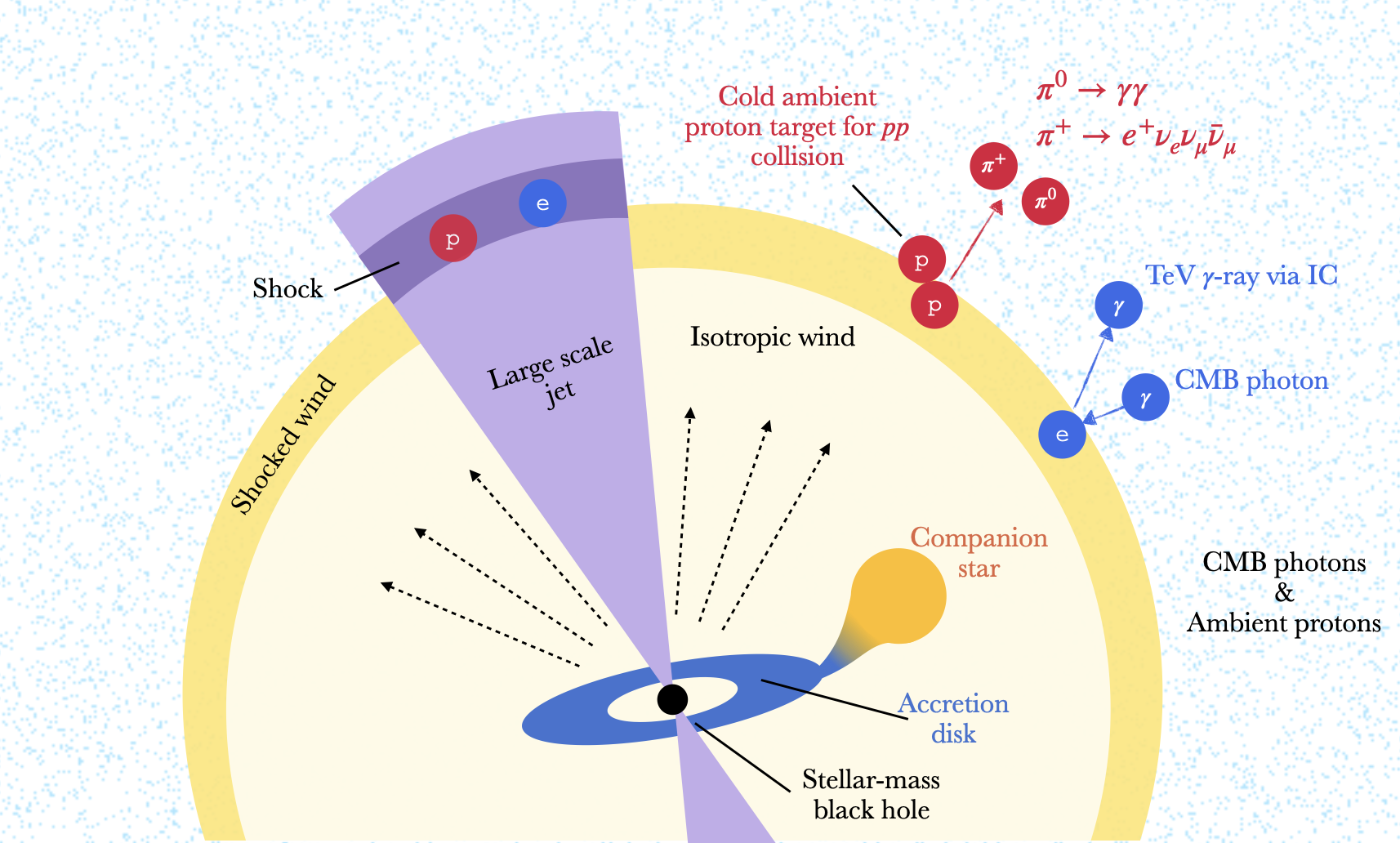}
\caption{Schematic of the model used in this study. Particles are accelerated via diffusive shock acceleration, both in the jet and in the isotropic wind. Electrons interact with the CMB photons and upscatter them to TeV energies via IC. }
\label{SchematicFig}
\end{figure}

To model the VHE emission, we assume that the acceleration of non-thermal particles occurs within a region of size $R$ with magnetic field $B$. For our study, acceleration occurs either at termination shocks in an isotropic wind or at shocks in large scale jets. The schematic for our model is shown in Fig. \ref{SchematicFig}. We will associate X-ray observations of SS 433 and V4641 Sgr to particle acceleration in jets. On the other hand, the extended VHE \gray emission regions from LHAASO will be assumed as indicative of a wind. The jet scenario in SS 433 is treated differently, as we assume that the acceleration region is the X-ray knot, of size $R\approx 8.1$ pc \cite{Safi-Harb_1997} and is located at a distance of $\approx 56$ pc from the compact object. The large difference between these two distances will have an implication for the cooling processes we will introduce later.

Following Ref. \cite{Kimura:2020acy}, we consider the X-ray knot as an acceleration site, with $V_{\rm adv} = 1.9\times 10^{9}\;{\rm cm \; s}^{-1}$. We take $R\approx 8.1$ pc based on the size of the e2 knot \cite{Safi-Harb_1997}. This size is smaller than the jet, of size $R_{\rm jet}=56$ pc. In this special setup, the adiabatic time becomes $t_{\rm adi}=R_{\rm jet}/V_{\rm adv}$ rather than $R/V_{\rm adv}$. For the extended source, we adopt a size of $R=100$ pc, which was also used in \cite{LHAASO:2024psv}. Here, the acceleration site would be at the termination shock. 

In the case of electrons with energies in the TeV energy range and above, efficient Compton scattering requires photons of energy $E_\gamma\lesssim 1\;$ eV, as more energetic photons have a lower cross section due the Klein-Nishina suppression~\cite{RevModPhys.42.237}. Hence, our main target for electrons is the CMB. The proton targets will be cold protons from the surrounding environment, as well as the CMB. 

We account for particle propagation by using a modified version of the \texttt{AM$^3$} package \cite{Klinger:2023zzv}, which solves the differential equation
\begin{equation}
\frac{\partial n_i}{\partial t} = -\frac{\partial}{\partial E_i}(\dot{E}_in_i)-\frac{n_i(E_i,t)}{t_{{\rm esc},i}}+\mathcal{Q}_i(E_i,t),
\label{TransportEq}
\end{equation}
where $n_i(E_i,t)$ is the differential number density of the particle species $i$, $\mathcal{Q}_i   $ is the differential injection rate, $t_{{\rm esc},i}$ is the escape time and $\dot{E}_i$ is the energy loss rate. 

For the escape process, we have the advection and diffusion timescales as
\begin{equation}
t_{\rm diff} = \frac{3eBR^2}{2c\eta E},
\end{equation}
and 
\begin{equation}
t_{\rm adv} = \frac{R}{V_{\rm adv}}.
\end{equation}
The parameter $\eta\geq 1$ is a dimensionless factor. We then define the escape time for the protons and electrons as $t_{\rm esc}= (t_{\rm diff}^{-1}+t_{\rm adv}^{-1})^{-1}$. The \grays and neutrinos are free-streaming, such that $t_{\rm esc} = R/c$.

The magnetic field strength within the acceleration regions is assumed to be $B\sim 10\;\mu{\rm G}$ for electrons to explain the X-ray emission (e.g., in SS 433 \cite{Reynoso:2019vrp} and V4641 Sgr \cite{Suzuki:2024rzc}).

Protons and electrons cool via synchrotron, with a characteristic timescale
\begin{equation}
t_{\rm syn} = \frac{6\pi m_i^4c^3}{\sigma_T B^2 E_i m_e^2},
\end{equation}
where $m_i$ is the particle mass, $m_e$ is the electron mass, and $\sigma_T$ is the Thomson cross section. Additionally, they are subject to adiabatic cooling with a timescale
$t_{\rm adi}\approx t_{\rm adv}$.
Protons may also cool via $pp$ interactions, with characteristic timescale
\begin{equation}
t_{pp} = (n_p \kappa_{pp}\sigma_{pp}c)^{-1},
\end{equation}
where $n_p$ is the target proton number density in the emission region, $\kappa_{pp}\approx 0.5$ is the inelasticity for $pp$ interactions, and $\sigma_{pp}$ is the $pp$ inelastic cross section. The optical depth for $pp$ interactions is
\begin{equation}
\tau_{pp}=n_p\sigma_{pp}R \sim 3\times 10^{-6} \left(\frac{n_p}{1\;{\rm cm}^{-3}}\right)\left(\frac{R}{10\;{\rm pc}}\right),
\end{equation}
suggesting that $pp$ interactions are quite inefficient. In this optically thin regime, the \gray and neutrino fluxes arising from $pp$ inelastic collisions are proportional to $n_p$.

The injection rate $\mathcal{Q}_i$ ($i=e,p$) is assumed as a powerlaw with an exponential cutoff
\begin{equation}
\mathcal{Q}_i(E_i) = C_i E_i^{-s}\exp(-E_i/E_{i,{\rm max}}),
\end{equation}
where $E_{i,{\rm max}}$ is the maximum particle energy and $C_i$ is a normalization factor chosen such that $\int dE_i E_i\mathcal{Q}_i = \epsilon_iL/\mathcal{V}$, where $L$ is the luminosity and $\mathcal{V}$ is the volume. For the volume, we take $\mathcal{V} = 4\pi R^3/3$. We use $\epsilon_i$ as the parameter that sets the normalization of the spectrum.  We assume that both electrons and protons are accelerated with the same powerlaw index $s$. 

The cutoff energy $E_{i,{\rm max}}$ is determined by balancing the acceleration timescale $t_{\rm acc}$ and loss timescale $t_{\rm loss}$. The acceleration timescale is given by
\begin{equation}
t_{\rm acc} \approx \frac{20\eta E_i}{3ceB\beta^2},
\end{equation}
where $\beta$ is the shock velocity. The loss rate is separated into $t_{\rm loss}^{-1}=t_{\rm esc}^{-1}+t_{\rm cool}^{-1}$, where $t_{\rm cool}$ is the cooling timescale. The cooling rate $t_{\rm cool}^{-1}$ is given by the sum of all relevant cooling rates. 
For electrons, the synchrotron cooling rate is the dominant energy loss mechanism; for protons, it is the advection timescale.
Hence, the maximum electron energy in this regime is given by
\begin{equation}
E_{e, {\rm max}} \approx \sqrt{\frac{9\pi e \beta^2m_e^2c^4}{10\eta \sigma_T B}}\simeq 7.3\times 10^3\;{\rm TeV}\; \left(\frac{B}{{\rm 10 \mu G}}\right)^{-1/2} \frac{\beta}{\eta^{1/2}}.
\end{equation}

For the microquasars being studied, we have $\beta\sim 0.1$ and $B\gtrsim 10\;{\rm \mu G}$, so electron energies typically won't exceed 700 TeV. 

During propagation, $\pi^\pm$ produced from $pp$ interactions promptly decay into neutrinos, without significant energy losses. The neutrino spectrum is then expected to follow the proton spectrum down to the minimum pion energy, below which the energy spectrum becomes constant. 
A similar spectral shape is seen for hadronic $\gamma$-rays at production, but reprocessing of \grays during propagation may distort the spectral shape.

An $E_e^{-s}$ injected electron spectrum would create an Inverse Compton power spectrum with index $(s-1)/2$ in the slow cooling regime, which is harder than the hadronic index for $s\geq 1$. This is true as long as $t_{\rm syn}>t_{\rm adv}$, which is the case up until the electron energy
\begin{equation}
E_{e,{\rm c}}=\frac{6\pi m_e^2 V_{\rm adv}c^3}{\sigma_T B^2 R}\approx 2.4\times 10^5\;{\rm GeV}\; \left(\frac{V_{\rm adv}}{1.9\times 10^9\;{\rm cm\; s}^{-1}}\right)\left(\frac{B}{10\;\mu{\rm G}}\right)^{-2}\left(\frac{R}{10\;{\rm pc}}\right)^{-1}.
\end{equation}
The average photon energy, in the Thomson regime, is $E_\gamma= 4E_e^2E_{\gamma, \rm CMB}/(3 m_e^2)$ \citep{Khangulyan:2023oaq}, where $E_{\gamma, \rm CMB}=7.1\times 10^{-4}\;{\rm eV}$ is the typical CMB photon energy. Hence, the photon spectrum will follow the slow cooling regime formula up to 
\begin{equation}
E_{\gamma, \rm c}\approx 2.1\times 10^5\;{\rm GeV}
\left(\frac{V_{\rm adv}}{1.9\times 10^9\;{\rm cm\; s}^{-1}}\right)^2\left(\frac{B}{10\;\mu{\rm G}}\right)^{-4}\left(\frac{R}{10\;{\rm pc}}\right)^{-2}.
\end{equation}
The spectral softening of the leptonic component at higher energies is influenced by three main effects: a transition to the fast cooling regime,  Inverse Compton in the Klein-Nishina regime,  and a spectral cutoff in the electron injection spectrum.

We obtain the densities $n_i$ by solving Eq. \eqref{TransportEq} until it reaches a steady state. We convert the densities to fluxes via
$\Phi_i = n_i \mathcal{V}/(4\pi d^2 t_{{\rm esc},i})$, where $d$ is the distance to the microquasar. For SS 433, we take $d=5.5$\; kpc \cite{Blundell:2004re}, V4641 Sgr $d=6.6$\; kpc \cite{Gaia2018}, and GRS 1915+105 $d=9.4$ \; kpc \cite{Reid:2023ksq}. As the emission from these sources are propagating through interstellar radiation fields and the CMB, the \gray fluxes are corrected with their associated attenuation factors $e^{-\tau}$ reported in \cite{Zhang:2024xkh}, where $\tau$ is the two-photon annihilation optical depth. While the value is both distance and energy-dependent, we note that for the three sources considered, attenuation is negligible below 10 TeV. However, $e^{-\tau}\approx 0.8-0.9$ around 100 TeV and $e^{-\tau}\approx 0.4-0.7$ around 1 PeV \cite{Zhang:2024xkh}. In the case of neutrinos, there is no attenuation due to propagation to Earth. However, neutrinos do oscillate, which modifies the 1:2:0 $\nu$+$\bar\nu$ flavor ratio from pion decay to 1:1:1 on Earth. 

\begin{table*}[t]
\centering
\begin{tabular}{lccccccccc}
   Source  & $L$ & $\beta$ &  $B$ & $R$ & $V_{\rm adv}$ & $\eta$ & $s$ & $\epsilon_e$ & $n_p$\\
     &  ${10^{39}\rm erg/s}$ & &  $\mu{\rm G}$ & pc & cm/s &  &  &  & ${\rm cm}^{-3}$\\
   \hline
   SS 433  &   &  & & & &  &  &  & \\ 
   
   Eastern lobe  & $2$& 0.26 & 25 & 8.1 & $1.9\times 10^9$ & 35 & 2 & $2\times 10^{-3}$ & 30\\
   extended source &  $1$ & 0.26 & 20 & 100 & $1.9\times 10^9$ & 50 & 2 & $10^{-4}$ & 2\\
   \hline
   V4641 Sgr  & &  & & & &  &  &  & \\  
   XRISM region & $6$ & 0.08 & 24 & 16 & $\beta c$ & 1.2 & 1.8 & $10^{-4}$ & 18\\
   HAWC region & $2$& 0.02 & 20 & 60 & $10^8$ & 2 & 1.9 & $10^{-4}$ & 2.8\\
   \hline
   GRS 1915+105 & $6$ & 0.05 & 15 & 30 & $\beta c$ & 2 & 2 & $10^{-4}$ & 22\\
   \hline
\end{tabular}
\caption{Model parameters used for each source in this study.}
\label{Table1}
\end{table*}

\subsection{SS 433}

X-ray observations from SS 433 point to the presence of X-ray lobes, denoted as {\em eastern} and {\em western}. 
Based on optical observations, the jet velocity is estimated to $\beta=0.26$ \cite{1979Natur.279..701A} and the kinetic energy of the jet is $L\approx 2\times 10^{39}{\rm \; erg\; s}^{-1}$. 

Interestingly, LHAASO found that the 25-100 TeV \gray component is dominated by two point-like sources associated with the east and west X-ray lobes \cite{LHAASO:2024psv}. However, above 100 TeV, the emission is explained better by an extended region. We thus consider the point-like sources to be associated with jets, while the extended emission is explained by an isotropic sub-relativistic wind. 

Following Ref. \cite{Kimura:2020acy}, we consider the X-ray knot as an acceleration site, with $V_{\rm adv} = 1.9\times 10^{9}\;{\rm cm \; s}^{-1}$ and $R\approx 8.1$ pc. As mentioned before, the size of the jet is $R_{\rm jet}=56$ pc and is much larger than the X-ray knot. Hence, the adiabatic time  $t_{\rm adi}=R_{\rm jet}/V_{\rm adv}$ rather than $R/V_{\rm adv}$. For the extended source, we adopt a size of $R=100$ pc, which was also used in \cite{LHAASO:2024psv}. Here, the acceleration site would be at the termination shock. The extended source should assume $\epsilon_e\lll 1$, because a magnetic field of order $\sim 10\mu{\rm G}$ would inevitably result in significant X-ray emission beyond the lobes, which is not observed.  

To reduce the number of parameters, we will assume the same values of $s, \beta, B$, and $V_{\rm adv}$ for both the wind and the jets. This leaves the wind luminosity for the extended source as a free parameter. We then use $\varepsilon_e$ and $n_p$ to modulate the flux normalizations and $\eta$ to adjust the maximum proton and electron energies. 

\begin{figure*}
    \begin{center}
    \includegraphics[width=0.8\columnwidth]{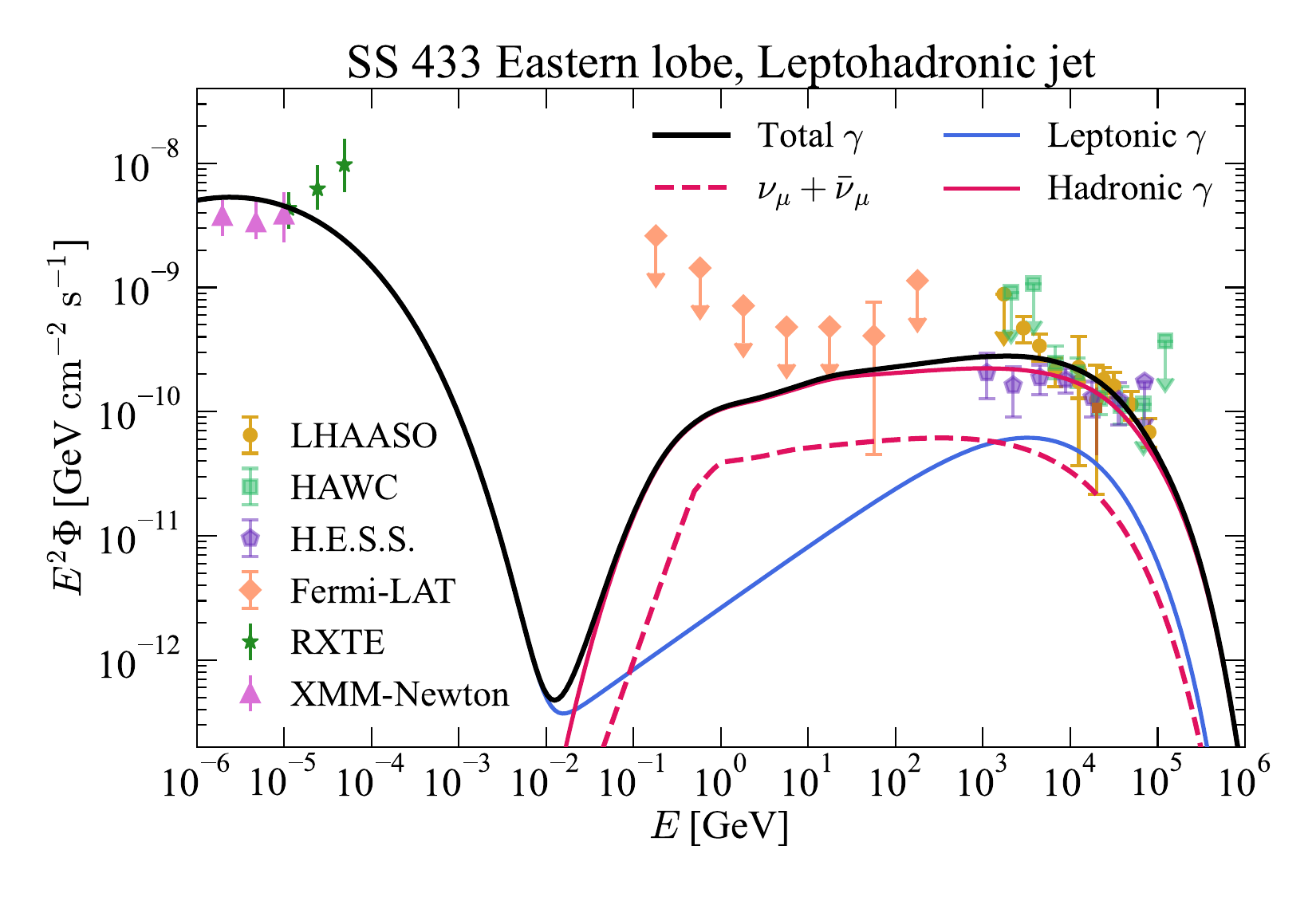}

    \caption{Multimessenger emission from the jet in the eastern lobe of SS 433. The total photon spectrum is shown as a thick black curve, and the leptonic (hadronic) component as a crimson (blue) thin curve. The dashed line corresponds to the predicted muon-neutrino spectrum. The eastern lobe electromagnetic measurements from XMM-Newton (magenta triangles) \cite{Brinkmann:2006zt}, RXTE (dark green stars) \cite{Safi-Harb_1999}, Fermi (orange diamonds, 95\% upper limits and $1\sigma$ bars) \cite{Fang:2020tcd}, HAWC (light green squares) \cite{HAWC:2024ysp}, H.E.S.S. (purple pentagons) \cite{HESS:2024rlh} and LHAASO  (golden circles) \cite{LHAASO:2024psv} are shown.}
    \label{SS433_Eastern}
    \end{center}
\end{figure*}

\begin{figure*}
    \begin{center}
    \includegraphics[width=0.8\columnwidth]{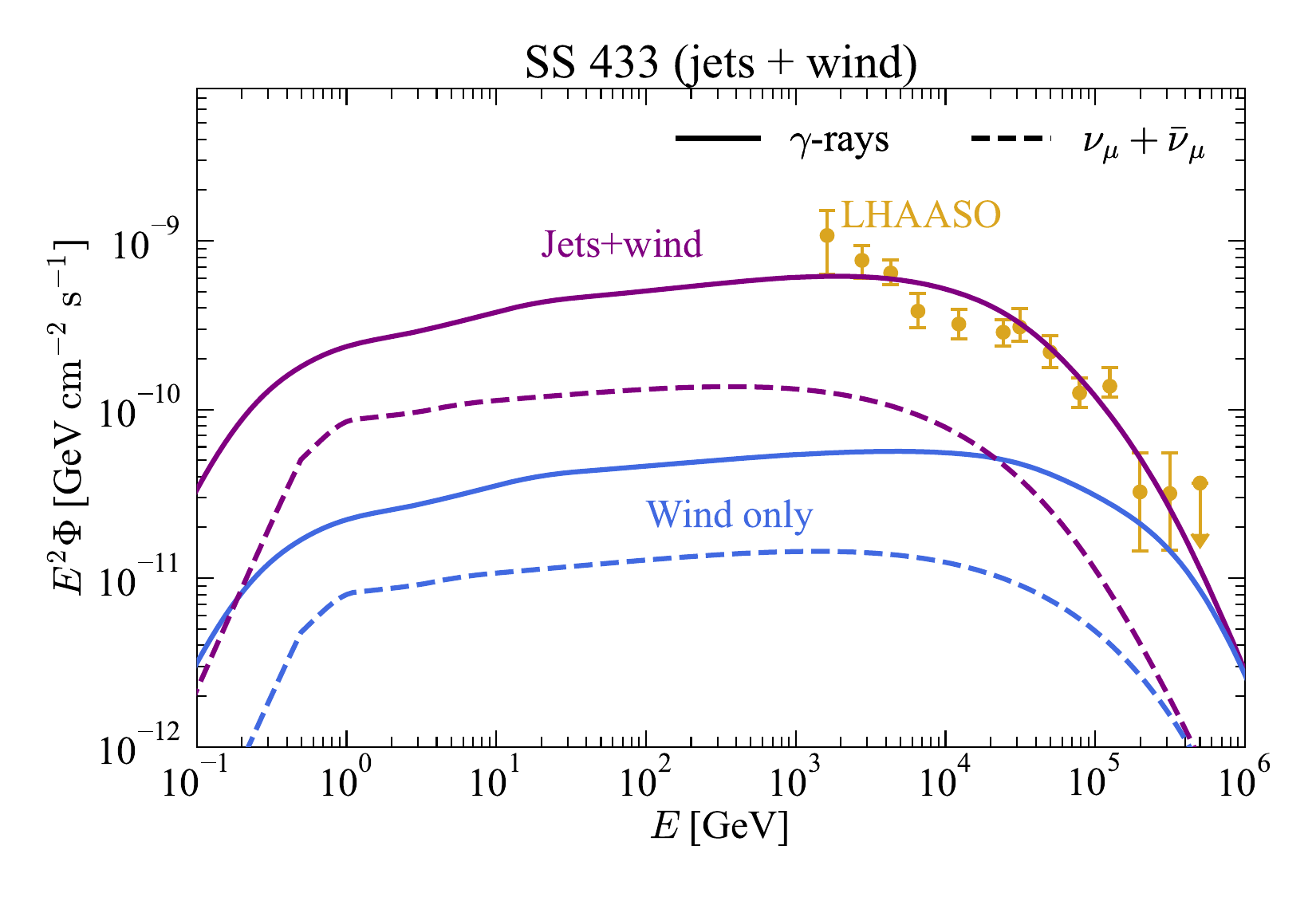}
    \caption{
\gray and neutrino emission associated with the SS 433 extended region identified by LHAASO. The purple (blue) curves correspond to the combined jet and wind (only wind) \grays. The dashed lines are the expected $\nu_\mu+\bar{\nu}_\mu$ fluxes. LHAASO  observations are shown as golden circles \cite{LHAASO:2024psv}.}
    \label{SS433_Total}
    \end{center}
\end{figure*}

\subsection{V4641 Sgr and GRS 1915+105}

V4641 Sgr and GRS 1915+105 are the two most significant LHAASO sources after SS 433, with significances of 8.1$\sigma$ and 6.1$\sigma$, respectively \cite{LHAASO:2024psv}. 

Assuming a hadronic origin for V4641 Sgr's VHE \gray emission around the X-ray region observed by XRISM ($\sim 20\;{\rm pc}$),  the lower limit on the magnetic field strength is constrained as $B\gtrsim 8\mu{\rm G}$ \cite{Suzuki:2024rzc}. On the other hand, HAWC data suggests an extended region of   $\sim 60$ pc \cite{Alfaro:2024cjd}. The different sizes of the emission regions affect our values of $t_{\rm esc}$ and $\tau_{pp}$. Hence, we consider $R=10$ pc and $R=60$ pc as two separate cases for our study.

For GRS 1915+105, GeV data from {\em Fermi}-LAT is also available \cite{Marti-Devesa:2024otf}. The X-ray data shows that the flux in the keV region exceeds $E^2\Phi > 10^{-9}\;$\ergcmsqs \cite{Titarchuk_2009}. This has been attributed to a thermal X-ray component, which is not part of this work. Hence, for this source we focus on the LHAASO and {\em Fermi} fluxes. The hadronic component helps to account for the relatively hard GeV spectrum, as a leptonic component would require a large injection time to account for the observed flux \cite{Marti-Devesa:2024otf}.

\begin{figure*}
    \begin{center}
    \includegraphics[width=0.88\linewidth]{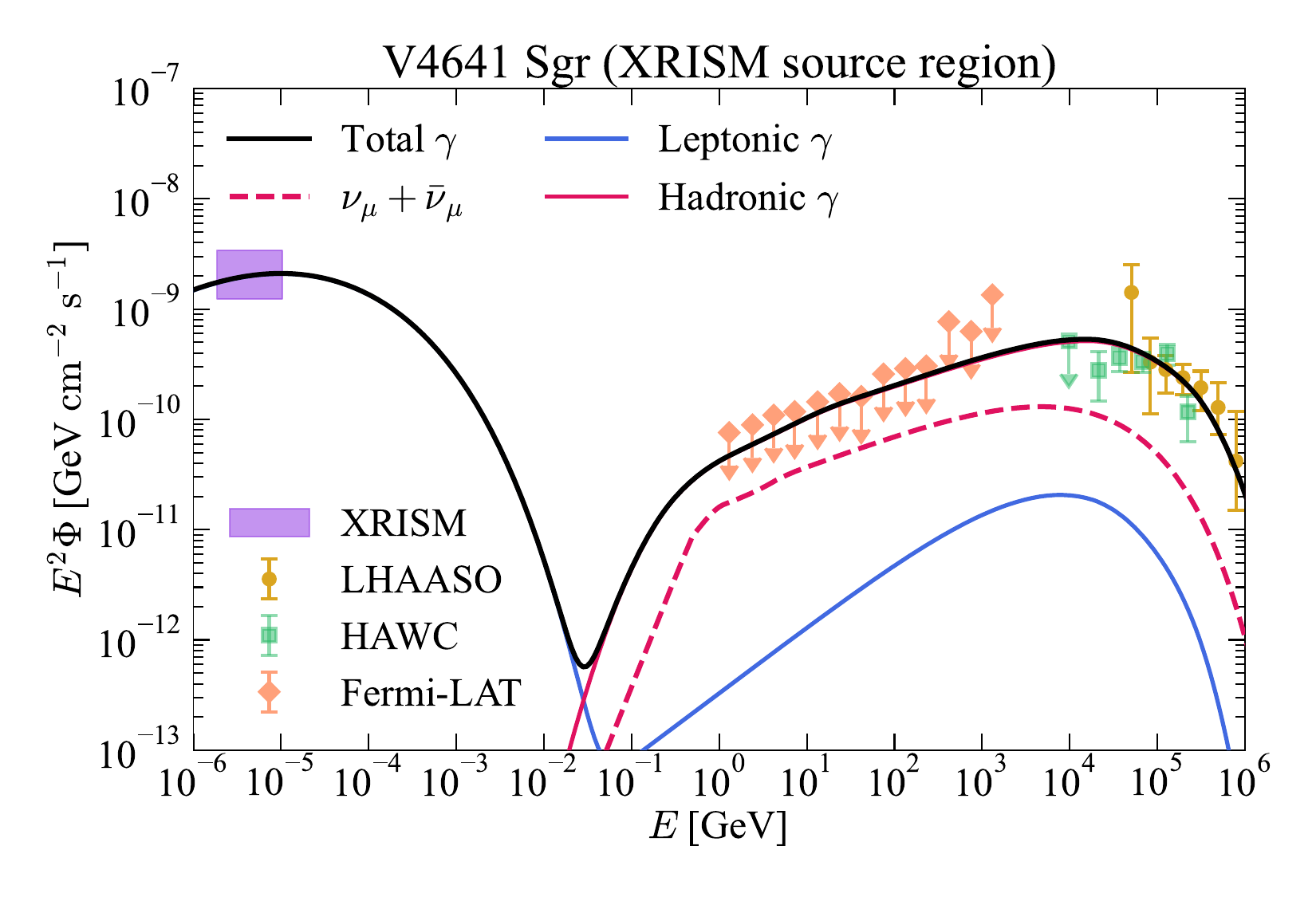}
    \includegraphics[width=0.88\linewidth]{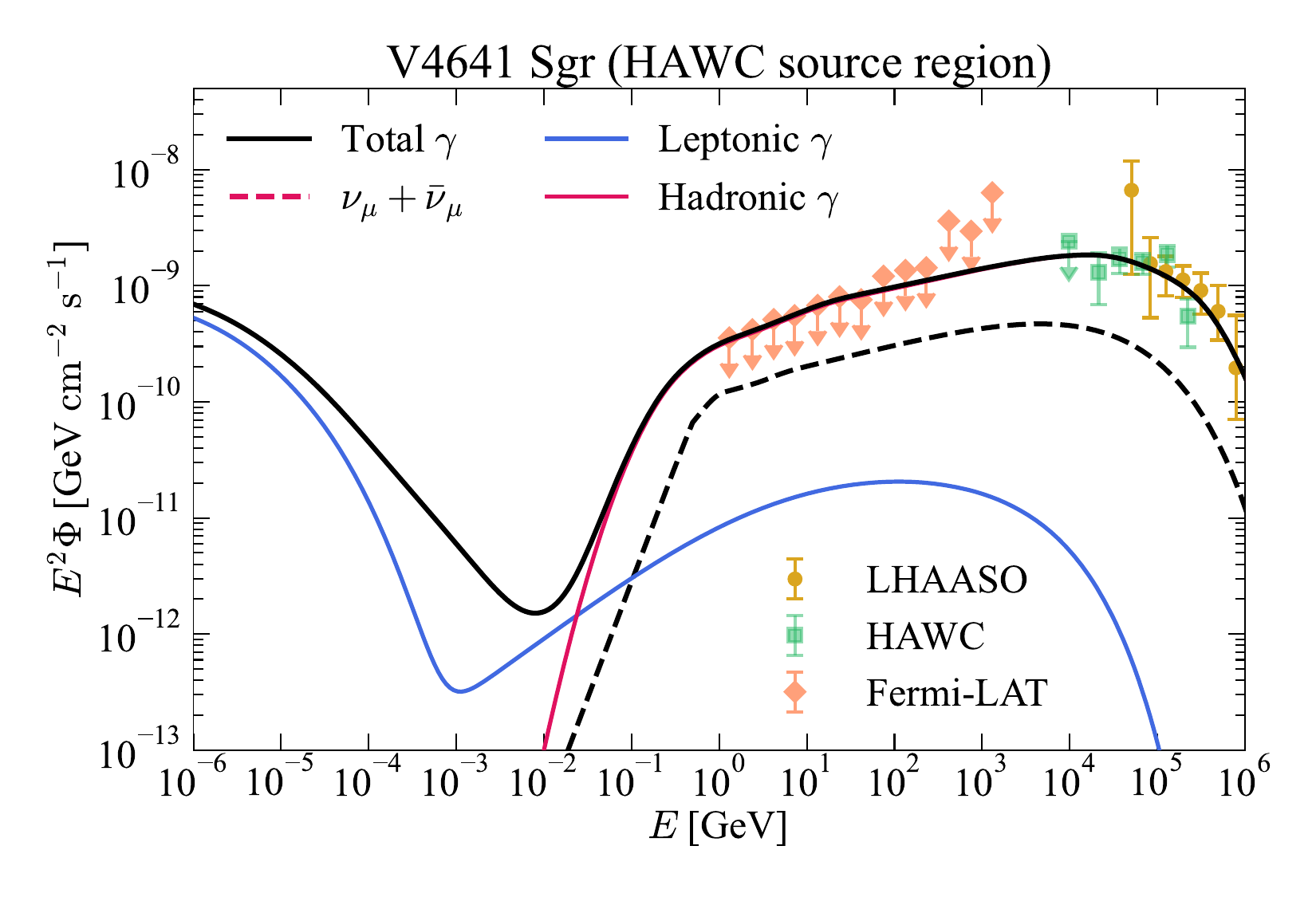}
    \caption{Top panel: Multimessenger emission from the XRISM X-ray source region in V4641 Sgr. Fermi-LAT \cite{Neronov:2024ycp}, HAWC \cite{HAWC:2024ysp} and LHAASO \cite{LHAASO:2024psv} are shown as orange diamonds, green squares and golden circles, respectively. XRISM X-ray band  shown as the shaded purple region \cite{Suzuki:2024rzc}. and HAWC data were scaled to the XRISM source region size, as done in \cite{Suzuki:2024rzc}. Bottom panel: Same as left panel, but assuming the HAWC source region instead.}
    \label{V4641-Spectra}
    \end{center}
\end{figure*}

\section{Results}
We apply the method outlined in Section \ref{sec:method} to calculate the multimessenger emission from SS433, V4641 Sgr and GRS 1915+105 and look for the parameter sets that are compatible with observed X-ray and \gray data. We then use those results to predict the neutrino spectra and detection prospects. 

Starting with SS 433, we look at the jet emission responsible for the eastern lobe emission separately from the isotropic hadronic wind creating the extended region. SS 433 is a well-studied source in both X-rays and \grays. The eastern lobe X-ray data is obtained from the XMM-Newton \cite{Brinkmann:2006zt} and RXTE \cite{Safi-Harb_1999}. The \gray observations from LHAASO are complemented by H.E.S.S. \cite{HESS:2024rlh} and HAWC \cite{HAWC:2024ysp}, as well as GeV measurements from {\em Fermi}-LAT \cite{Fermi-LAT:2019yla}.

We show our results in Figures~\ref{SS433_Eastern} and \ref{SS433_Total}. We adopted $s=2$, $B=20\;\mu{\rm G}$, $\beta=0.26$, and $V_{\rm adv}=1.9\times 10^9$ cm s$^{-1}$ as common parameters for the wind and jet components.
Starting with the leptohadronic jet scenario, shown in Figure~\ref{SS433_Eastern}, we assume a proton density of $n_p=30\;{\rm cm}^{-3}$. The leptonic component accounts for the X-ray observations from XMM-Newton and RXTE. It also contributes to the 1 TeV -- 100 TeV $\gamma$-rays in the eastern lobe. For the GeV range, both the leptonic and hadronic components by themselves are below the {\em Fermi} upper limits. Nevertheless, the leptonic component is well below the 100 MeV -- 100 GeV {\em Fermi} upper limits, making the hadronic component more likely to explain the GeV data. Further GeV measurements would provide further insight in confirming the presence of a hadronic component.

Our results vary slightly from those shown in \cite{Kimura:2020acy}. In their study, only one $>$ 10 TeV data point was available from HAWC data \cite{Fang:2020tcd} at $\approx 10^{-10}$ GeV cm$^{-2}$ s$^{-1}$. Currently, significantly more data is available above 1 TeV, which allows for a more robust study of the high-energy emission. As we adopted the same values of $L,\beta, V_{\rm adv}$ and $R$, we compare their hadronic model (scenario A in \cite{Kimura:2020acy}) against our hadronic component. The main difference is our use of a larger $n_p$ (30 cm$^{-3}$ instead of 10 cm$^{-3}$) and $\eta$ (25 instead of 2). The change in the normalization is to accommodate the $\sim 50-100$ TeV tail of the \gray emission, which is at the level of $\gtrsim 10^{-10}$ GeV cm$^{-2}$ s$^{-1}$, while our updated value of $\eta$ lowers $E_{p, {\rm max}}$ and reproduces the tail from LHAASO data. Our values of $\epsilon_e$ and $B$ are comparable since the X-ray data from the leptonic component provides less flexibility for these two parameters.

\begin{figure}
    \begin{center}
    \includegraphics[width=0.8\textwidth]{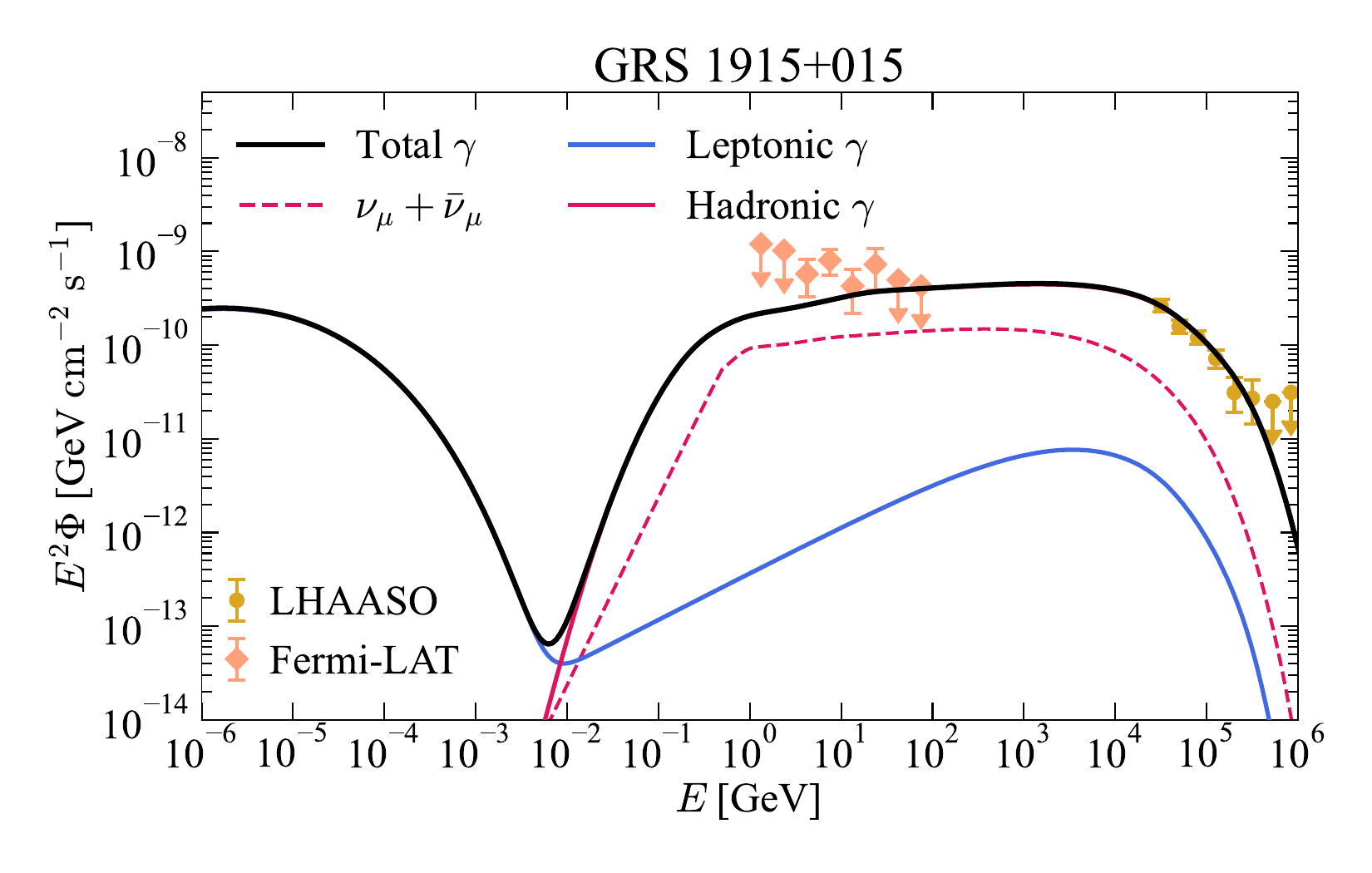}
    \caption{Multimessenger emission from GRS 1915+105. {\em Fermi}-LAT \cite{Marti-Devesa:2024otf} and LHAASO \cite{LHAASO:2024psv} are shown as orange diamonds and golden circles, respectively.}
    \label{GRSSpectra}
    \end{center}
\end{figure}
The flux from the wind is shown in Fig.~\ref{SS433_Total}, as well as the combined contributions from both the jets and the wind. For simplicity, we assume that the east and west lobes have identical fluxes, given that the  corresponding  fluxes reported by LHAASO are similar \cite{LHAASO:2024psv}.
The much larger region in the case of the wind component has a reduced $pp$ cooling rate, which allows for a larger maximum proton energy. This allows the wind component to explain the high-energy \gray tail above 100 GeV. We find that the wind luminosity is half the jet's luminosity, which can explain the LHAASO data above 100 TeV.

The multimessenger emission of V4641 Sgr is shown in the top (bottom) panel of Fig. \ref{V4641-Spectra} for the XRISM (HAWC) observed region. In the X-ray identified region, of size $R=16$ pc, we found that very efficient acceleration ($\eta=1.5$) was needed to obtain the $>100$ TeV $\gamma$-rays.
In addition, the proton spectral index is restricted to $s\lesssim 2$, with $s=1.8$ being consistent with the {\em Fermi} GeV bounds. A softer spectrum requires a significant reduction in $n_p$ for pionic \grays to remain consistent with the GeV constraints. In that case, the leptonic component would have to provide the majority of the $>10$ TeV $\gamma$-rays, including LHAASO's. This generally leads to X-ray fluxes that are much larger than XRISM observations. 

\begin{figure*}
    \begin{center}
    \includegraphics[width=0.8\linewidth]{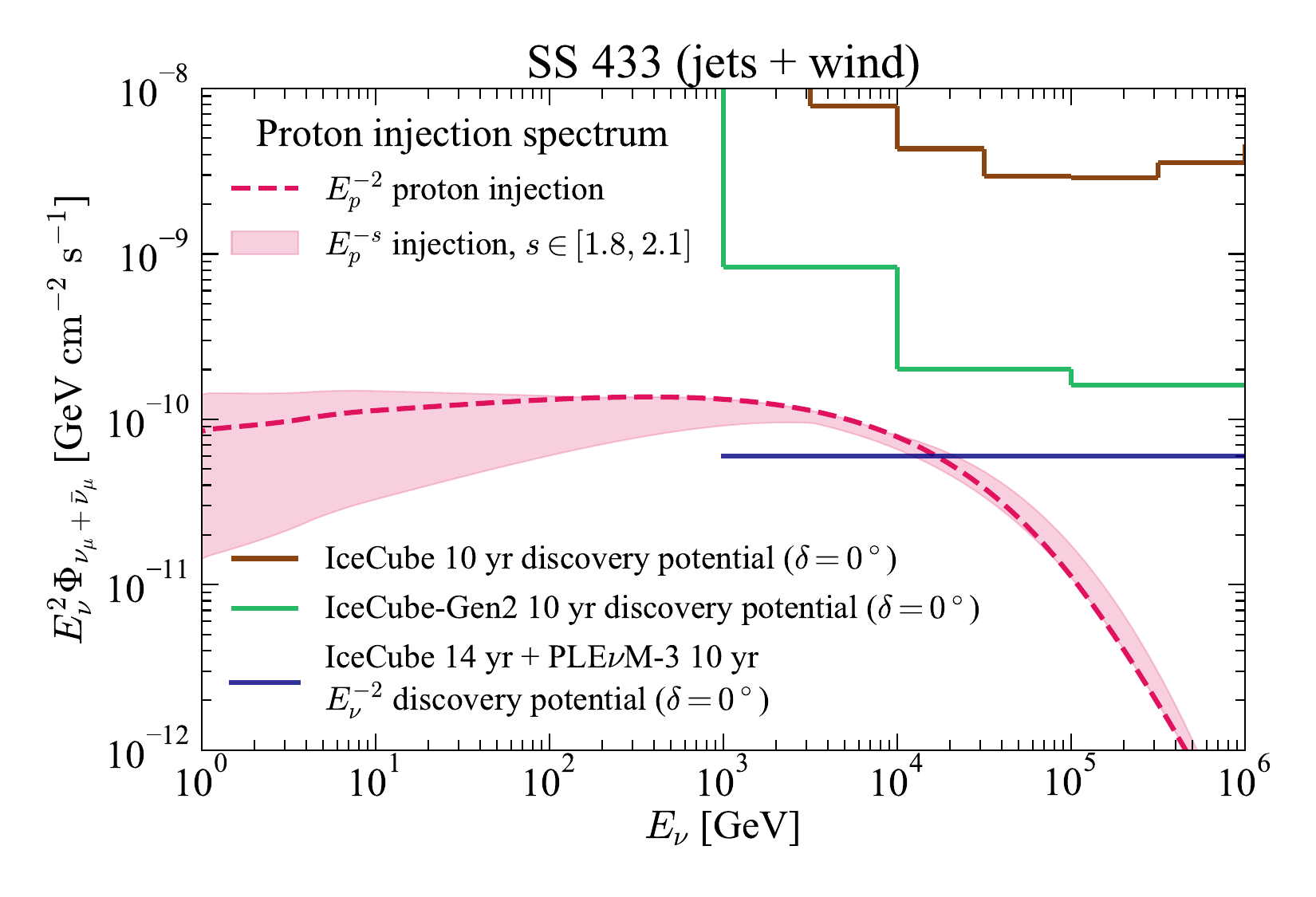}
    \includegraphics[width=0.8\linewidth]{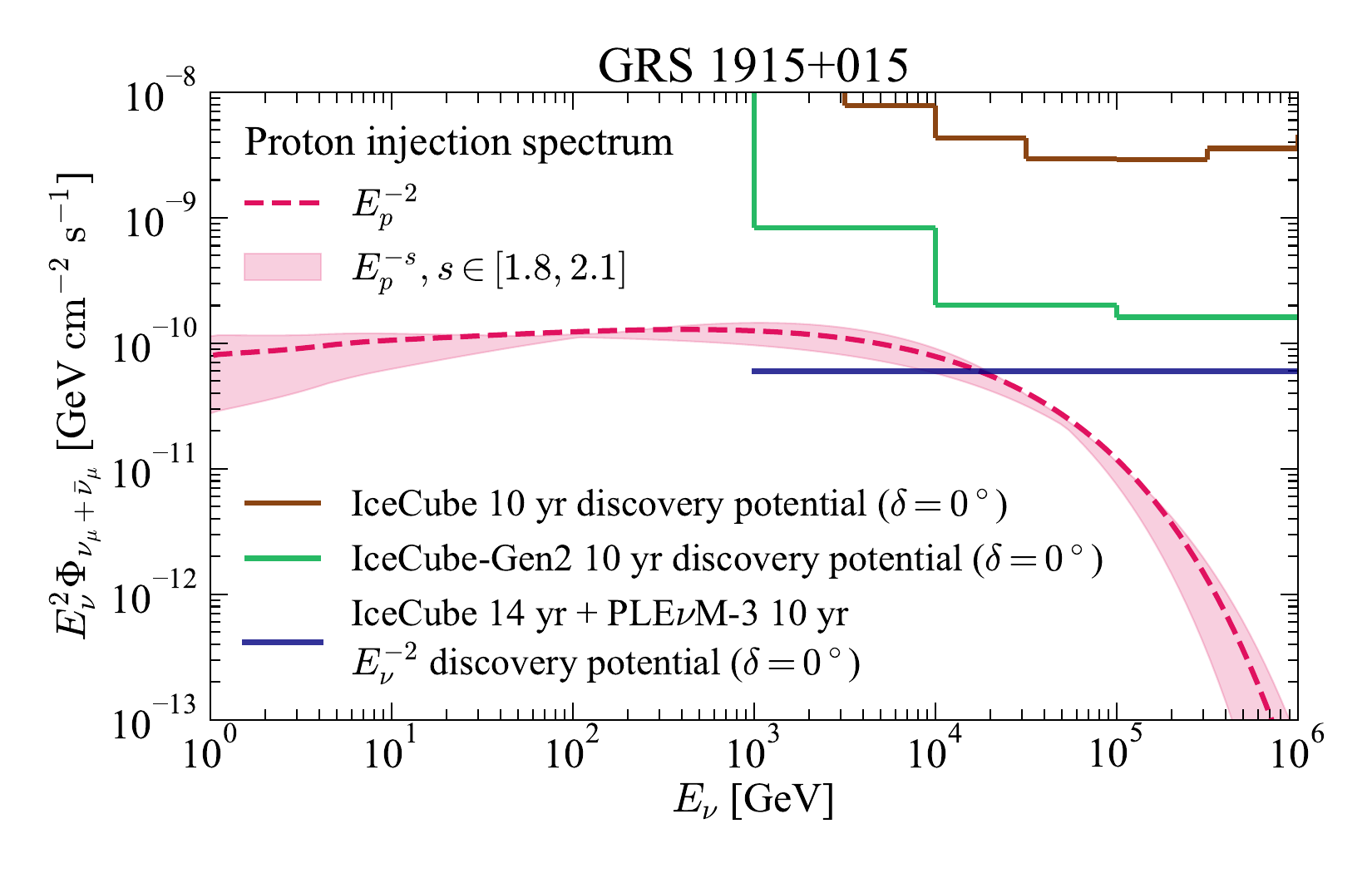}
    \caption{Top panel: $\nu_\mu+\bar\nu_\mu$ emission from SS 433, combining the jets and wind contributions. Bottom panel: Same as the top panel, for GRS 1915+105. The 10 year discovery potential for a point source at the horizon for IceCube \cite{Glauch:2021kke} and IceCube-Gen2 \cite{IceCube-Gen2:2020qha} are shown by brown lines and green lines, respectively. The $E_\nu^{-2}$ discovery potential at the horizon for 10 years of PLE$\nu$M-3 is shown as a purple line \cite{Schumacher:2025qca}.}
    \label{NeutrinoSpectrum1}
    \end{center}
\end{figure*}

When applying our model to the HAWC's observed region, of size $R=60$ pc, the escape time is  longer. This means that more IC and $pp$ scatterings may take place, allowing for larger \gray and neutrino fluxes. Similar to the XRISM region scenario, we used $s=1.9$ to allow for larger values of $n_p$ while staying below {\em Fermi} upper limits. Likewise, efficient acceleration ($\eta=1.2$) and $V_{\rm adv}=10^8\;{\rm cm \;s}^{-1}$ were needed to allow $E_{p,{\rm max}}\approx 5$ PeV and explain the high-energy \gray tail. We show the resultant fluxes in the bottom panel of Fig.~\ref{V4641-Spectra}. Within our leptohadronic model, we chose $\epsilon_e=10^{-4}$, where the X-ray emission remains below the XRISM measurement and peaks below 1 keV.

Finally, we show the GRS 1915+105 hadronic component in the right panel of Fig. \ref{GRSSpectra}. Our model explains the LHAASO data whilst remaining consistent with {\em Fermi} observations. We find that a softer proton injection spectrum explaining the LHAASO spectrum would be incompatible with the {\em Fermi} upper bounds above 10 GeV. The observed X-ray flux is at the level of $E^2\Phi\sim 10^{-6}$ \ergcmsqs, originating from the vicinity of the black hole \cite{Titarchuk_2009}. For completeness, we show that the synchrotron emission from the hadronic emission is several orders of magnitude below the observed X-ray flux, meaning that X-ray data does not restrict our model. 

For the corresponding neutrino emission, we assess the detectability of these sources by current and future neutrino detectors. SS 433 and GRS 1915+105 are located in the Northern Sky, at declinations $\delta=4.98^\circ$ and $\delta= 10.9^\circ$, respectively. This makes them good candidates for detection via muon tracks in IceCube. On the other hand, V4641 Sgr is in the Southern sky, with $\delta= - 25.7^\circ$. 
For this source, we compare the projected fluxes with KM3NeT sensitivity given that it offers a better sensitivity to the Southern sky \cite{KM3NeT:2024uhg}. In addition, we consider the case of a combined detector measurement via the monitoring network PLE$\nu$M \cite{Schumacher:2025qca}, specifically PLE$\nu$M-3 which combines P-ONE~\cite{P-ONE:2020ljt}, IceCube-Gen2~\cite{IceCube:2015uxl}, TRIDENT~\cite{TRIDENT:2022hql}, NEON~\cite{Zhang:2024slv}, and HUNT~\cite{Huang:2023R8}.

The projected muon neutrino flux from SS 433 (jets + wind) and from GRS 1915+105 are shown in Fig.~\ref{NeutrinoSpectrum1}. Given their vicinity to the horizon, we show the IceCube and IceCube-Gen2 10-year point source discovery potentials at $\delta = 0^\circ$, as reported in \cite{Glauch:2021kke} and \cite{IceCube-Gen2:2020qha}, respectively. We also included the $E^{-2}$ point source discovery potential from 10 years of PLE$\nu$M-3 and 14 years of IceCube. For both sources, we also studied the neutrino emission for $E^{-s}$ spectra, with $s\in [1.8,2.1]$. When varying the source parameters, we used the parameters in Table~\ref{Table1} and varied $s$, $\eta$, $\epsilon_e$ and $n_p$ to accommodate the 1 GeV -- 100 TeV \gray data. We vary $\eta$ to adjust the high-energy cutoff and and vary $\epsilon_e$ and $n_p$ to adjust the normalizations. 

In both sources, we see that the fluxes are not large enough to reach the IceCube-Gen2 differential discovery potential. However, the 1 TeV -- 10 TeV flux is above the $E^{-2}$ discovery potential in a combined detector configuration. In SS 433, a harder spectrum of $E^{-1.8}$ is allowed because the {\em Fermi} data point at $\approx 50$ GeV has a large error bar. In GRS 1915+105, we could not accommodate a spectrum harder than $E^{-1.8}$ without increasing the normalization to values that would overshoot the LHAASO data.

\begin{figure}
\centering
    \includegraphics[width=0.75\columnwidth]{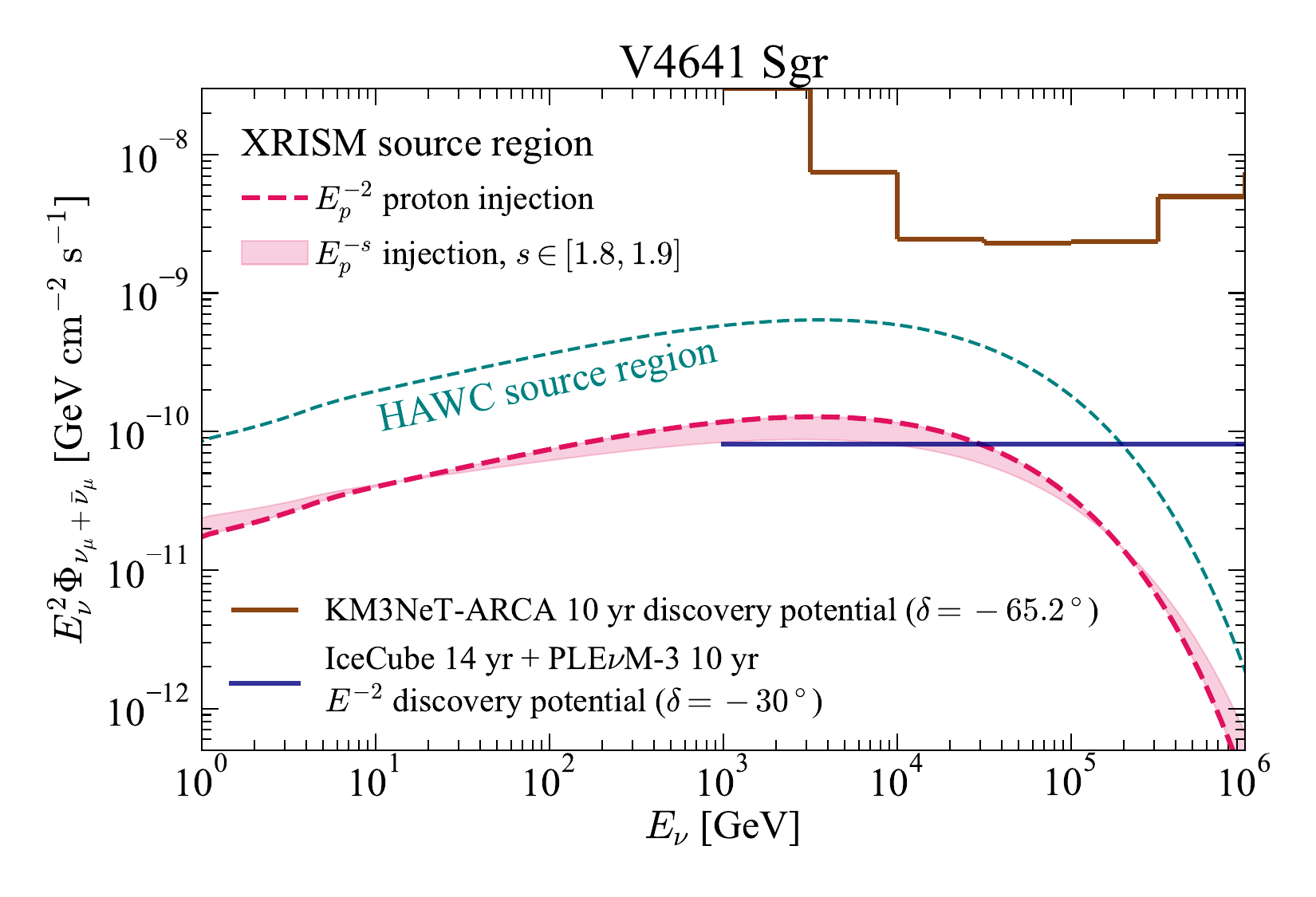}
    \caption{$\nu_\mu+\bar\nu_\mu$ emission from V4641 Sgr. The dashed red curve corresponds to the XRISM observed region, while the teal curve represents the HAWC observed region. The 10 year discovery potential at $\delta=-65.2^\circ$ for KM3NeT-ARCA is shown as brown lines \cite{KM3NeT:2024uhg}.  $E_\nu^{-2}$ discovery potential for a point source at $\delta=-30^\circ$ for 10 years of PLE$\nu$M-3 is shown as a purple line \cite{Schumacher:2025qca}.}
    \label{NeutrinoSpectrum2}
\end{figure}

Given the location of V4641 Sgr, we used the KM3NeT-ARCA discovery potentials at $\delta = -65.2^\circ$, reported in \cite{KM3NeT:2024uhg}. 
We use this declination as a representative for a Southern sky source, as a differential sensitivity was not reported at a declination closer to V4641 Sgr. For PLE$\nu$M-3, we used the $5\sigma$ discovery potential at $\delta=-30^\circ$ \cite{Schumacher:2025qca}. Whether we choose the models describing the HAWC or XRISM observed regions, the fluxes do not reach the KM3NeT sensitivity, as shown in Fig. \ref{NeutrinoSpectrum2}.
Our parameter choice for the region observed by HAWC provides a  promising scenario, with a flux that is above the $E^{-2}$ discovery curve for energies below $\approx 300$\; TeV. Combining the next-generation neutrino detectors provide an avenue to confirm the hadronic origin of the $>$ 100 TeV \grays, or provide constraints on either $\varepsilon_p$ or $n_p$ in the event of a non-detection. 

\section{Summary \& Discussions}
In this work, we studied the multimessenger emission from three of the VHE \gray LHAASO-identified microquasars, using a phenomenological model.
We considered both leptonic and hadronic models to explain the VHE \gray emission. Using the electromagnetic data to find a compatible model parameter set, we also highlighted the neutrino detection prospects with current and multiple upcoming neutrino telescopes.
The \gray emission tends to disfavor proton and electron spectra softer than $E^{-2}$ due to {\em Fermi} GeV data. We relied on the hadronic component to account for the $> 100$\; TeV \grays, with an ambient proton density of $n_p\sim1-10\;{\rm cm}^{-3}$. It is difficult to achieve an electron maximum energy $E_{e,{\rm max}}\gtrsim \, 300\;{\rm TeV}$ for the leptonic component to significantly contribute to the $> 100\;{\rm TeV}$ \gray flux. Strong synchrotron cooling due to $B\sim 10\mu {\rm G}$ limits $E_{e,{\rm max}}$ and we cannot significantly lower $B$ without being in tension with X-ray observations. 

In the LHAASO Collaboration's study, a lepto-hadronic model of SS 433 allows for the leptonic component to explain the \gray emission, up to $\approx$ 10 TeV, with an $E^{-2.4}$ electron spectrum and a super-exponential cutoff $\propto \exp(-E_e^2/E_{e,{\rm max}}^2)$ with $E_{e,{\rm max}}=200$\; TeV~\cite{LHAASO:2024psv}. 
The electrons were assumed to be accelerated at jet terminations and required an electron luminosity above 1 TeV of $\simeq 6\times 10^{35}\;{\rm erg\; s}^{-1}$. This amounts to a total electron luminosity (above $m_e$) of $\sim 10^{38}\;{\rm erg\; s}^{-1}$, implying $\epsilon_e\sim 0.1$. 
Within our model, we could find good agreement with the LHAASO \gray data by assuming $s=2.3$ and $\epsilon_e=0.1$. However, such a high value for $\epsilon_e$ would overshoot the X-ray data due to excessive synchrotron radiation.

Our approach relied on $pp$ interactions to account for the hadronic $\gamma$-rays. However, it is possible for additional photon fields to initiate $p\gamma$ interactions. For 100 TeV protons, the minimum photon energy for $p\gamma$ interactions to occur is 
\begin{equation}
E_{\gamma,{\rm min}} = \frac{2m_\pi m_p+m_\pi^2}{4E_p} \approx 700 \left(\frac{100\;{\rm TeV}}{E_p}\right)\;{\rm eV}.
\end{equation}

Hence, background X-ray and $\gamma$-rays would allow additional $\pi^0$ production, but the associated photon densities from synchrotron and IC emissions within the extended regions are too low to achieve an efficient pion production. It is possible for particle acceleration to occur in the corona, near the black hole, where X-ray targets are more abundant. However, this is a more likely scenario when the \gray emission region is more compact, such as the case of Cygnus X-1 (see e.g., \cite{Fang:2024wmf}).

We find that the neutrino emission from the three sources may be detectable by combining the next-generation of neutrino telescopes.  Among the 3 sources, our models indicate that V4641 Sgr is the most likely source to be detected, provided that the particle acceleration site has a size comparable to the HAWC observed region. 

Here, we compared our results with the point source sensitivities for muon track. Inclusion of cascades in such searches will enhance the sensitivity, for instance in IceCube as it offers a better sensitivity in the Southern sky~\cite{Abbasi:2021ryj}. Furthermore, new analysis techniques aiming at identifying starting tracks will additionally improved the sensitivity of neutrino source searches~\cite{IceCube:2025zyb}. It is worth noting that the efforts to combine these data streams~\cite{IceCube:2023gtp} would enhance the sensitivity of these searches and provide the best accessible sensitivity in current detectors. 

Our results show that sub-relativistic outflows can explain the $>100$ TeV LHAASO \grays. We highlight the importance of multimessenger observations to obtain a deeper understanding of these accelerators. We find that the GeV measurements from {\em Fermi} - LAT are shown to be crucial to the reject purely leptonic scenario and help narrowing down the spectral index of hadronic contributions. Neutrinos serve as the determinant observables to identify the hadronic origin of \grays.

\section*{Acknowledgements}
We would like to thank Juanan Aguilar for useful discussions. AK and BZ are supported by NASA grant 80NSSC23M0104.
J.C. acknowledges the support from Nevada Center for Astrophysics through the NCfA fellowship and NASA award 80NSSC23M0104.

\bibliographystyle{elsarticle-num} 
\bibliography{main.bib}

\end{document}